\documentclass[10pt,twocolumn]{article}

\usepackage[
  letterpaper,
  left=0.75in,
  right=0.75in,
  top=1.0in,
  bottom=1.0in,
  columnsep=0.25in
]{geometry}

\usepackage[T1]{fontenc}
\usepackage{times}
\usepackage{microtype}
\usepackage{ragged2e}
\usepackage{natbib}

\usepackage{amsmath,amssymb,amsthm,mathtools}
\usepackage{graphicx}
\usepackage{float}
\usepackage{subfigure}
\usepackage{booktabs}
\usepackage{hyperref}
\usepackage[capitalize,noabbrev]{cleveref}
\usepackage{placeins}
\usepackage{dblfloatfix}

\usepackage{titlesec}
\titleformat{\section}{\large\bfseries}{\thesection}{0.5em}{}
\titleformat{\subsection}{\normalsize\bfseries}{\thesubsection}{0.5em}{}
\titleformat{\subsubsection}{\normalsize\scshape}{\thesubsubsection}{0.5em}{}

\usepackage{fancyhdr}
\fancypagestyle{plain}{
  \fancyhf{}
  \fancyfoot[C]{\small\thepage}
  
}

\begin{document}

\twocolumn[
\begin{center}
\rule{\linewidth}{1pt}

\vspace{1.4em}
{\Large\bfseries
Inspectable Neural Markov Models for Non-Stationary Time Series}
\vspace{0.5em}

\rule{\linewidth}{1pt}

\vspace{0.9em}

{\bfseries Jan Rovirosa \quad Jesse Schmolze}\\
{\small University of Wisconsin -- Madison, Madison, WI, USA}
\end{center}

\vspace{1.0em}
]

\thispagestyle{plain}

\begin{center}
\begin{minipage}{0.8\linewidth}
{\centering\bfseries Abstract\par}
\vspace{0.5em}
\small\normalfont
\justifying
\setlength{\parindent}{0em}
\setlength{\parskip}{0pt}

Modeling non-stationary stochastic systems requires balancing the representational capacity of deep learning with the structural transparency of classical probabilistic models. Markov transition matrices provide such a framework, but traditional frequency-based estimation collapses at high resolutions due to data sparsity. We propose a hybrid approach that parameterizes the manifold of stochastic matrices through a neural network, enabling estimation of time-inhomogeneous Markov chains in sparse-data regimes, and use financial markets as a testbed to investigate the Markov state variable as a critical inductive bias. We show that conditioning on realized volatility produces a more internally consistent Markovian structure than return-based states, achieving a $5.6\%$ reduction in Chapman-Kolmogorov discrepancy and superior held-out likelihood in 9 of 10 assets. Unlike black-box sequence models, our approach generates explicit matrices amenable to direct geometric analysis, surfacing structural findings such as the universal homogenization of transition probabilities under high-volatility regimes.

\end{minipage}
\end{center}

\vspace{0.75em}

\section{Introduction}
\label{sec:introduction}
The rise of deep generative modeling has provided unprecedented predictive power, yet these models often operate as ``black boxes'', lacking the structural constraints necessary for rigorous analysis in high-stakes domains. In contrast, classical Markov transition matrices provide a transparent, structured framework where system evolution is represented as an explicit stochastic law. However, these matrices are notoriously difficult to estimate in high-dimensional or sparse data regimes. Discretizing a continuous process into $n$ states creates $n^2$ possible transitions; at high resolutions, empirical frequency-based estimators collapse into sparse, unreliable matrices.

We propose a hybrid framework that integrates deep learning with classical Markovian structure, yielding explicit transition matrices that expose the model’s internal logic. By replacing traditional frequency-based counting with neural parameterization, we maintain structural transparency without sacrificing the representational power of deep neural networks. We treat the transition matrix as a function of observable features, using a neural network to map the current context to the manifold of valid stochastic matrices. This approach allows the model to ``fill in'' the gaps of the state space by learning the underlying functional form of the transitions. For portfolio managers holding $k$ assets, this provides a granular $k \times N$ matrix (where $N$ is the number of bins) of transition probabilities across the portfolio, offering a direct view of the predictive distribution which can be key for risk management and asset allocation. 

A critical design choice in this structured framework is the definition of the Markov state. We demonstrate that the mathematical consistency of the model depends heavily on this choice. While returns are a common state variable, they are often serially uncorrelated. We show that conditioning the framework on realized volatility produces a more internally consistent Markovian structure. Specifically, the volatility-state model achieves a significant reduction in Chapman-Kolmogorov discrepancy, suggesting that it better satisfies the structural assumptions of the Markov property than traditional return-based approaches.

\paragraph{Contributions.}
\begin{itemize}
    \item We develop a hybrid generative framework that integrates the flexibility of deep neural networks with the structural constraints of time-inhomogeneous Markov chains. By parameterizing the manifold of stochastic matrices directly, we maintain high-dimensional representational power while enabling the use of rigorous, interpretable diagnostics that are typically unavailable in black-box sequence models.
    
    \item We demonstrate that the choice of state variable acts as a critical inductive bias within this structured framework. Specifically, we show that conditioning on realized volatility rather than returns results in a more internally consistent Markovian structure, evidenced by a $\sim 5.6\%$ lower average Chapman-Kolmogorov discrepancy (Section~\ref{sec:results_consistency}).

    \item We validate our framework on a 24-year sample of ten US bank stocks, showing that the volatility-state model achieves superior held-out likelihood in 9 of 10 assets and uncovers universal structural insights (such as the homogenization of transition probabilities during high-volatility regimes) directly from the model's internal geometry.
\end{itemize}

\section{Related Work}
\label{sec:related_work}

\paragraph{Markov Modeling in Finance.} 
Markov transition matrices have a long history as transparent models for regime shifts in economic systems. \citet{hamilton1989new} demonstrated that non-stationary macroeconomic dynamics can be captured through regime-dependent evolution governed by a discrete-state Markov law. In financial settings, \citet{lando2002analyzing} applied these tools to credit rating migrations, while \citet{mettle2022analysis} used time-inhomogeneous chains to model exchange rate volatility. While our work shares this motivation, it departs from classical approaches by replacing frequency-based estimation (counting) with neural parameterization. This allows the model to remain robust in high-resolution settings where empirical counts typically collapse due to data sparsity.

\paragraph{Neural Time-Series Forecasting.} 
From an estimation perspective, our task is a conditional density estimation problem. Early neural approaches used Mixture Density Networks (MDNs) \citep{bishop1994mixture} to parameterize output distributions. Modern architectures like DeepAR \cite{salinas2020deepar} and other neural sequence models \cite{lim2021time} have become the standard for probabilistic forecasting. However, these models are structurally different from our approach: they typically represent dynamics through latent recurrent states or parameters at the observation level. They do not produce an explicit stochastic matrix on a fixed state space, meaning they cannot be subjected to the classical Markovian diagnostics (like entropy or consistency checks) that we utilize here.

\paragraph{Operator-First Models.} 
Our framework is a ``Deep Markov'' \citep{krishnan2017structured} extension that bridges classical theory with deep learning. A key precursor is the Input-Output HMM \cite{bengio1995input}, which conditions latent-state transitions on observed sequences. Our approach differs because our state variable is an explicit discretization of an observed process (returns or volatility) rather than a latent one. Our primary object of study is the transition matrix itself. This ``operator-first'' perspective is compatible with the views expressed in distributional reinforcement learning \cite{bellemare2017distributional}, which treats probability distributions as primary objects. By using a network to parameterize the manifold of stochastic matrices, we maintain structural rigor while gaining the representational power of deep learning.

\section{Framework}\label{sec:framework}

\subsection{State Space and Discretization}
\label{sec:state_space}

Let $\{P_t\}_{t \geq 0}$ denote daily prices and define the log-return as $r_t := \log P_t - \log P_{t-1}$. To analyze the system's dynamics, we map these continuous values into a discrete state space $\mathcal{S} = \{s_1, \ldots, s_n\}$ using quantile-based bins calculated from the training data.

The primary design choice is the \textit{state variable} $X_t \in \mathcal{S}$. We compare two approaches:
\begin{itemize}
    \item \textbf{Return-state:} $X_t = \varphi(r_t)$. This defines the state based on the asset's immediate performance (the return).
    \item \textbf{Volatility-state:} $X_t = \varphi(\sigma_t)$, where $\sigma_t$ is a rolling estimate of realized volatility. This defines the state based on the current level of market uncertainty.
\end{itemize}

We use a high resolution ($N = 35, 45, 55$ bins) to capture the nuances of the distributions. However, this creates a major sparsity problem: there are $n^2 = 3,025$ possible state-to-state transitions. In our dataset, over $99\%$ of these potential transitions have fewer than five observations in the training set, making traditional counting methods unreliable. We resolve this using the neural parameterization described below.

\subsection{Feature-Conditioned Transition Matrices}
\label{sec:matrix_def}

Let $\mathbf{F}_t \in \mathbb{R}^d$ be a vector of observable features available at time $t$ (e.g., macro indicators and lagged returns). We model the system as a time-inhomogeneous Markov chain where the transition laws change based on $\mathbf{F}_t$. For a prediction horizon $h \geq 1$, the transition matrix $A_t^{(h)}$ is an $n \times n$ stochastic matrix:
\begin{equation}
    A_t^{(h)}(i,j) := P(X_{t+h} = s_j \mid X_t = s_i, \mathbf{F}_t).
    \label{eq:matrix}
\end{equation}
Each row of this matrix is a valid probability distribution. Because these matrices share a consistent state space $\mathcal{S}$, we can compose them over time to check for mathematical consistency.

\subsection{Neural Parameterization (Deep Markov Construction)}
\label{sec:neural_param}

To overcome data sparsity, we use a neural network $f_\theta^{(h)}$ to estimate the transition probabilities. Instead of counting how many times the system moved from $s_i$ to $s_j$, the network learns a functional mapping from the current context to the next-step distribution:
\begin{equation}
    \widehat{A}_t^{(h)}(i, \cdot) := f_\theta^{(h)}(s_i, \mathbf{F}_t).
    \label{eq:row_estimator}
\end{equation}

To construct the complete $n \times n$ matrix for a specific timestep, we fix the features $\mathbf{F}_t$ and evaluate the network $n$ times, once for every possible starting state $s_i$. This allows the model to ``fill in'' transitions that may never have been observed in the training data by generalizing from similar states and feature sets. The result is a mathematically valid stochastic matrix at every timestep that is fully inspectable.

\paragraph{Architecture.} The model is a Multi-Layer Perceptron (MLP) that takes a one-hot encoded state $\mathbf{e}(X_t)$ and the feature vector $\mathbf{F}_t$ as input. It uses hidden layers of widths $64 \to 128 \to 256 \to 128 \to 64$ with GELU activations and a final softmax layer to ensure the output is a valid probability distribution. 

\paragraph{Feature-only baseline.} To isolate the value of the Markov state, we compare our model against a baseline that ignores $X_t$ and relies solely on the features $\mathbf{F}_t$. In this model, every row in the transition matrix is identical, representing a system with no state-dependence.

\subsection{Volatility-State Conditioning}
\label{sec:vol_conditioning}

\paragraph{Motivation.} While returns are the most intuitive state variable, they are often uncorrelated over time, making them poor predictors of future behavior. In contrast, realized volatility is highly persistent: high-volatility periods tend to last for days or weeks. A Markov state based on volatility ($\sigma_t$) therefore captures the ``regime'' of the system, a property known as volatility clustering and central to the GARCH family of models \citep{engle1982autoregressive, bollerslev1986generalized}, more effectively than a state based on the return.

\paragraph{Definition.} We calculate rolling realized volatility over a window $w$ as:
\begin{equation}
    \sigma_t^{(w)} := \sqrt{\frac{1}{w}\sum_{k=0}^{w-1}(r_{t-k} - \bar{r}_t^{(w)})^2}.
    \label{eq:sigma}
\end{equation}
We use a 21-day window ($w=21$) as our default, as this timeframe best captures the persistence of volatility in equity markets.

\subsection{Structural Diagnostics}
\label{sec:diagnostics}

Because our model outputs an explicit matrix rather than a hidden internal state, we can apply classical diagnostics to ``see'' what the model has learned.

\paragraph{Row Heterogeneity.} This measures how much the next-step distribution changes depending on the current state. It is the average Total Variation (TV) distance between all pairs of rows in $\widehat{A}_t$. If heterogeneity is zero, the Markov state is irrelevant; higher values indicate that the current state strongly influences the system's future.

\paragraph{Row Entropy.} This measures the model's uncertainty. We calculate the Shannon entropy for each row and average them. High entropy means the model predicts a nearly uniform distribution (high uncertainty), while low entropy suggests the model has identified a clear path for the next transition.

\paragraph{Dobrushin Coefficient.} To assess the mixing properties of the learned dynamics, we compute the Dobrushin coefficient $\delta(\widehat{A}_t)$, defined as the maximum Total Variation distance between any two rows:
\begin{equation}
    \delta(\widehat{A}_t) := \max_{i,j} \text{TV}(\widehat{A}_t(i,\cdot), \widehat{A}_t(j,\cdot)).
    \label{eq:dobrushin}
\end{equation}
In Markov chain theory, $\delta < 1$ implies that the matrix is a strict contraction, ensuring the system converges toward a unique stationary distribution. This metric allows us to monitor how ``contractive'' or ``persistent'' different market regimes are.

\paragraph{Chapman-Kolmogorov (CK) Consistency.} A rigorous test of the first-order Markov assumption is the Chapman-Kolmogorov identity. We compare the model's direct $h$-step prediction, $\widehat{A}_t^{(h)}$, against the composition of $h$ sequential one-step predictions:
\begin{equation}
    \widehat{A}_t^{(1:h)} := \widehat{A}_t^{(1)} \widehat{A}_{t+1}^{(1)} \cdots \widehat{A}_{t+h-1}^{(1)}.
    \label{eq:ck_identity}
\end{equation}
We measure the discrepancy between these two matrices using the row-averaged KL divergence. In a perfectly Markovian system, this discrepancy would be zero, in reality, spikes in this ``CK error'' serve as an interpretable diagnostic for structural breaks where the local Markov approximation fails.

\section{Experimental Setup}\label{sec:experimental_setup}

\subsection{Dataset and Assets}
\label{sec:dataset}

We evaluate the framework on a universe of ten US bank stocks:
JPMorgan Chase (JPM), Citigroup (C), Wells Fargo (WFC), Goldman
Sachs (GS), Morgan Stanley (MS), PNC Financial (PNC), US Bancorp
(USB), Fifth Third Bancorp (FITB), M\&T Bank (MTB), and Bank of
America (BAC).  Daily adjusted closing prices span June 4, 1999,
through December 29, 2023, covering 6{,}183 trading days per asset
and encompassing multiple distinct market regimes: the dot-com
bust and post-9/11 drawdown of 2000--2002, the 2008--2009 global
financial crisis, the prolonged low-volatility expansion of
2013--2019, the COVID-19 shock of 2020, and the rate-driven
stress of 2022--2023.

All assets are processed independently: bin edges, feature
statistics, and preprocessing parameters are fit on each asset's
own training segment.  We apply a fixed chronological
train/validation/test split of $70\%/15\%/15\%$ along the time
axis, ensuring strict temporal ordering with no look-ahead bias
at any stage.

\subsection{Feature Set}
\label{sec:features}

A rich macro and cross-asset feature vector $\mathbf{F}_t \in
\mathbb{R}^d$ is constructed for each trading day $t$.  Features
fall into four groups.

\paragraph{Market and macro indicators.}
Equity volatility and sentiment: CBOE VIX, VXN, SKEW index, and
equity put/call ratio.  Fixed income: 2-year, 5-year, and 10-year
US Treasury yields, the 2s10s term spread, Fed Funds futures
(implied rate expectations), and TIPS breakeven inflation.
Credit: CDX Investment Grade and High Yield indices, Bloomberg
US HY Option-Adjusted Spread.  Cross-asset: DXY dollar index,
EUR/USD, USD/JPY, WTI crude, gold, and copper spot prices.
Equity indices: SPX, XLF, and Russell 2000 daily returns.
Financial conditions: Bloomberg US Financial Conditions Index (BFCI).

\paragraph{Cross-asset lagged returns.}
For each of the ten assets, we include lagged returns at horizons
of 1, 2, and 5 trading days, yielding $10 \times 3 = 30$ lagged
return features.  These are engineered from raw daily returns and
capture short-term cross-sectional momentum and mean-reversion
dynamics within the banking sector.

\paragraph{Per-asset rolling volatility.}
For each asset, we include its own rolling realized volatility
at window lengths of 5 and 21 trading days (Equation~\ref{eq:sigma}),
giving 2 volatility features per asset.  When the vol-state model
is used, $\sigma_t^{(21)}$ additionally determines the Markov
state $X_t$; it remains available as a feature to all models
regardless.

\paragraph{Preprocessing.}
All features are standardized using training-set mean and standard
deviation only.  Missing values arising from data unavailability
at the start of the sample are filled by carrying forward the
most recently available observation; features requiring more than
21 days of extrapolation are excluded.  The final feature
dimension is $d = 68$ after exclusions.

\subsection{Models}
\label{sec:models}

We evaluate five models, all sharing the identical MLP architecture
described in Section~\ref{sec:neural_param} and differing only
in their Markov state variable:

\begin{itemize}
    \item \textbf{\texttt{state\_free}}: Non-markovian baseline, essentially a feature-only $\mathbf{F}_t$ model.  Operator rows are identical at every $t$ by construction. 

    \item \textbf{\texttt{state\_cond\_return}}: Return-state
          conditioning; $X_t = \varphi(r_t)$.

    \item \textbf{\texttt{state\_cond\_vol\_w5}},
          \textbf{\texttt{state\_cond\_vol\_w10}},
          \textbf{\texttt{state\_cond\_vol\_w21}}: Volatility-state conditioning at window lengths $w \in \{5, 10, 21\}$; $X_t = \varphi(\sigma_t^{(w)})$.
\end{itemize}

All models are trained independently for each asset.  To assess
result stability, every configuration is run with three random
seeds ($\{42, 7, 123\}$); reported metrics are means over seeds
unless otherwise noted, with error bars indicating one standard
deviation across seeds.

\subsection{Evaluation Protocol}
\label{sec:evaluation}

\paragraph{Prediction targets and horizons.}
For each asset we construct horizon-specific datasets for
$h \in \{1, 2, 3, 5\}$ trading days.  The prediction target at
horizon $h$ is the discretized $h$-day forward return
$Y_t^{(h)} = \varphi(R_t^{(h)})$, where
$R_t^{(h)} = (P_{t+h} - P_t)/P_t$, discretized into $N$ output
bins.  We evaluate across resolutions $N \in \{35, 45, 55\}$.
For operator diagnostics (row heterogeneity, row entropy,
Dobrushin coefficient, Chapman-Kolmogorov consistency), we
use the state-to-state setting $Y_t^{(h)} = X_{t+h}$, which
yields square $N \times N$ operators on the shared state space
$\mathcal{S}$.

\paragraph{Precision metric.}
The main metric used to evaluate how accurate the model is when choosing the correct bin for tomorrow's change in price, is the test-set negative
log-likelihood, which is benchmarked against the marginal baseline:
\begin{equation}
    \Delta\mathrm{NLL}
    \;:=\;
    \mathrm{NLL}_{\mathrm{marginal}} - \mathrm{NLL}_{\mathrm{model}},
    \label{eq:delta_nll}
\end{equation}
where $\mathrm{NLL}_{\mathrm{marginal}}$ is the NLL of the
empirical marginal distribution of $Y_t^{(h)}$ on the test set.
Negative values indicate the model beats the marginal baseline;
more negative is better.  We report $\Delta\mathrm{NLL}$ rather
than raw NLL to enable fair comparison across assets and
resolutions whose marginal entropies differ.

\paragraph{Structural Diagnostics.}
To evaluate the learned dynamics, we compute row heterogeneity, row entropy, and the Dobrushin coefficient at every test-period timestep from the $N=55$ transition matrices. We additionally assess the internal mathematical consistency of the framework by measuring the Chapman-Kolmogorov (CK) discrepancy, the KL divergence between direct $h$-step predictions and composed one-step transitions. For the volatility-state model, this diagnostic reveals a $\sim 5.6\%$ lower CK discrepancy on average compared to the return-state alternative, indicating a more coherent Markovian structure. Lastly, cross-asset synchrony is assessed by computing the pairwise Pearson correlation of row entropy time series across all ten assets, using a 21-day moving average to reduce noise.

\section{Results}
\label{sec:results}

We organize the results around three questions. First, does volatility-state conditioning produce better predictions than return-state conditioning and the marginal baseline across the portfolio? Second, does it produce a more internally consistent Markov structure? Third, what does the explicit transition matrix reveal about the underlying dynamics that a black-box predictor would not?

\subsection{Predictive Performance}
\label{sec:results_predictive}

\paragraph{Pooled comparison at $h=1$, $N=55$.}
Table~\ref{tab:pooled_nll} reports mean test NLL and $\Delta$NLL
relative to the marginal baseline, pooled across all ten assets and three seeds. The \texttt{vol\_w21} model achieves the best pooled $\Delta\mathrm{NLL} = -0.0226$, outperforming \texttt{state\_cond\_return} ($-0.0131$), \texttt{state\_free} ($-0.0143$), and the macro-only ablation ($-0.0135$). All models beat the marginal baseline on average; the vol-state advantage over return-state conditioning amounts to $0.0095$ nats pooled, a modest but consistent improvement that holds across the portfolio.

\begin{table}[h]
\centering
\caption{Mean test NLL and $\Delta$NLL vs.\ marginal baseline
($h=1$, $N=55$, pooled over 10 assets and 3 seeds).
Lower NLL and more negative $\Delta$NLL are better.}
\label{tab:pooled_nll}
\begin{tabular}{lcc}
\hline
\textbf{Model} & \textbf{NLL} & \boldmath$\Delta$\textbf{NLL} \\
\hline
\texttt{state\_cond\_vol\_w21} & 3.9847 & $-0.0226$ \\
\texttt{state\_cond\_vol\_w10} & 3.9864 & $-0.0209$ \\
\texttt{state\_free}           & 3.9930 & $-0.0143$ \\
\texttt{state\_cond\_macro}    & 3.9938 & $-0.0135$ \\
\texttt{state\_cond\_return}   & 3.9942 & $-0.0131$ \\
\texttt{state\_cond\_vol\_w5}  & 3.9955 & $-0.0119$ \\
\hline
\end{tabular}
\end{table}

\begin{figure}[t]
    \centering
    \includegraphics[width=1.05\linewidth]{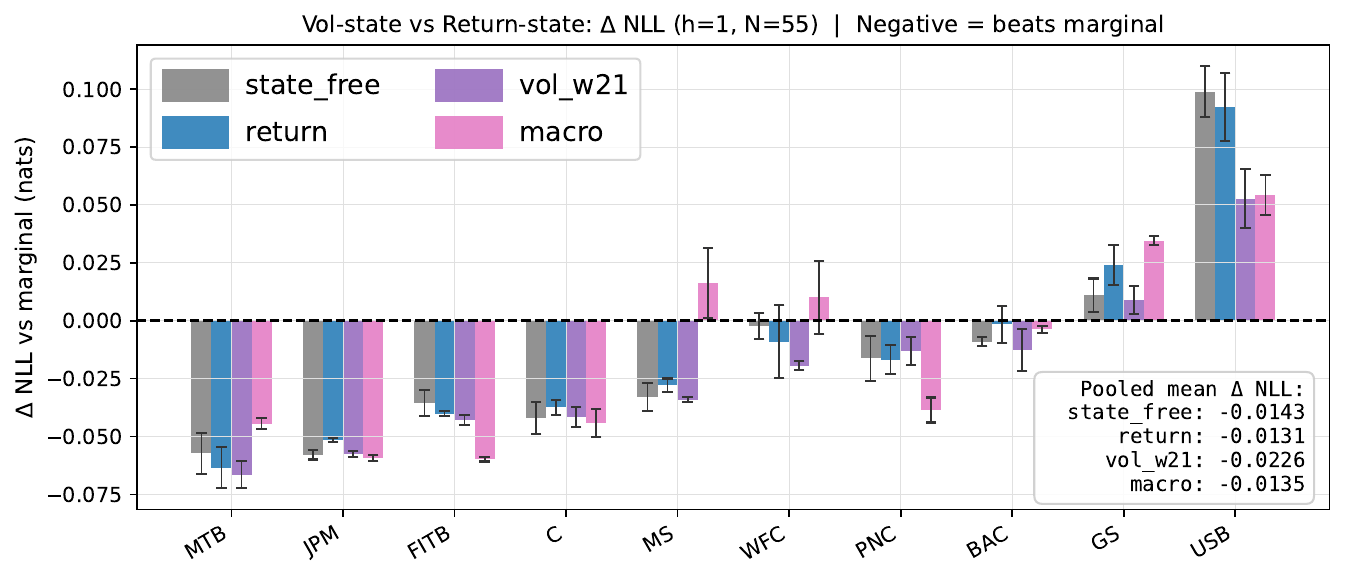}
    \caption{Per-asset $\Delta$NLL vs.\ marginal baseline
    ($h=1$, $N=55$). Negative values indicate the model beats
    the marginal. Error bars show one standard deviation across
    three seeds. \texttt{vol\_w21} achieves the best pooled
    $\Delta\mathrm{NLL} = -0.0226$ and beats
    \texttt{state\_cond\_return} in 9 of 10 assets.}
    \label{fig:headtohead}
\end{figure}

\begin{figure}[t]
    \centering
    \includegraphics[width=0.72\linewidth]{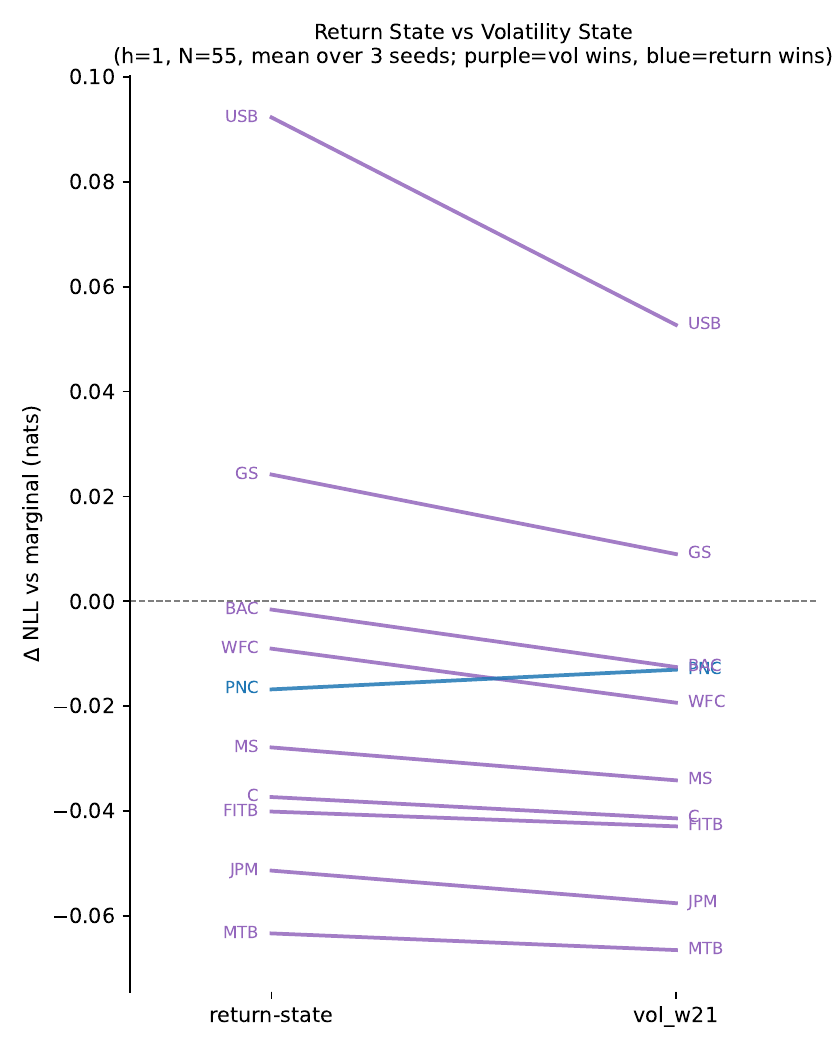}
    \caption{Paired $\Delta$NLL comparison of
    \texttt{state\_cond\_return} vs.\ \texttt{vol\_w21} per
    asset. Purple lines indicate assets where vol-state wins;
    blue lines indicate return-state wins. Vol-state
    conditioning improves NLL in 9 of 10 assets.}
    \label{fig:slopechart}
\end{figure}

\paragraph{Per-asset comparison.}
Figure~\ref{fig:headtohead} shows per-asset $\Delta$NLL for the
four main models, and Figure~\ref{fig:slopechart} pairs
\texttt{state\_cond\_return} directly against \texttt{vol\_w21}.
The vol-state model wins in 9 of 10 assets, with PNC the only
exception. WFC, BAC, and USB show the largest improvements;
JPM, MS, C, FITB, MTB, and GS show smaller but consistent
gains. USB is notable in that all conditioning types fail to
beat the marginal baseline at $h=1$, suggesting idiosyncratic
dynamics that none of the state variables capture at daily
frequency.

\begin{figure}[t]
    \centering
    \includegraphics[width=1\linewidth]{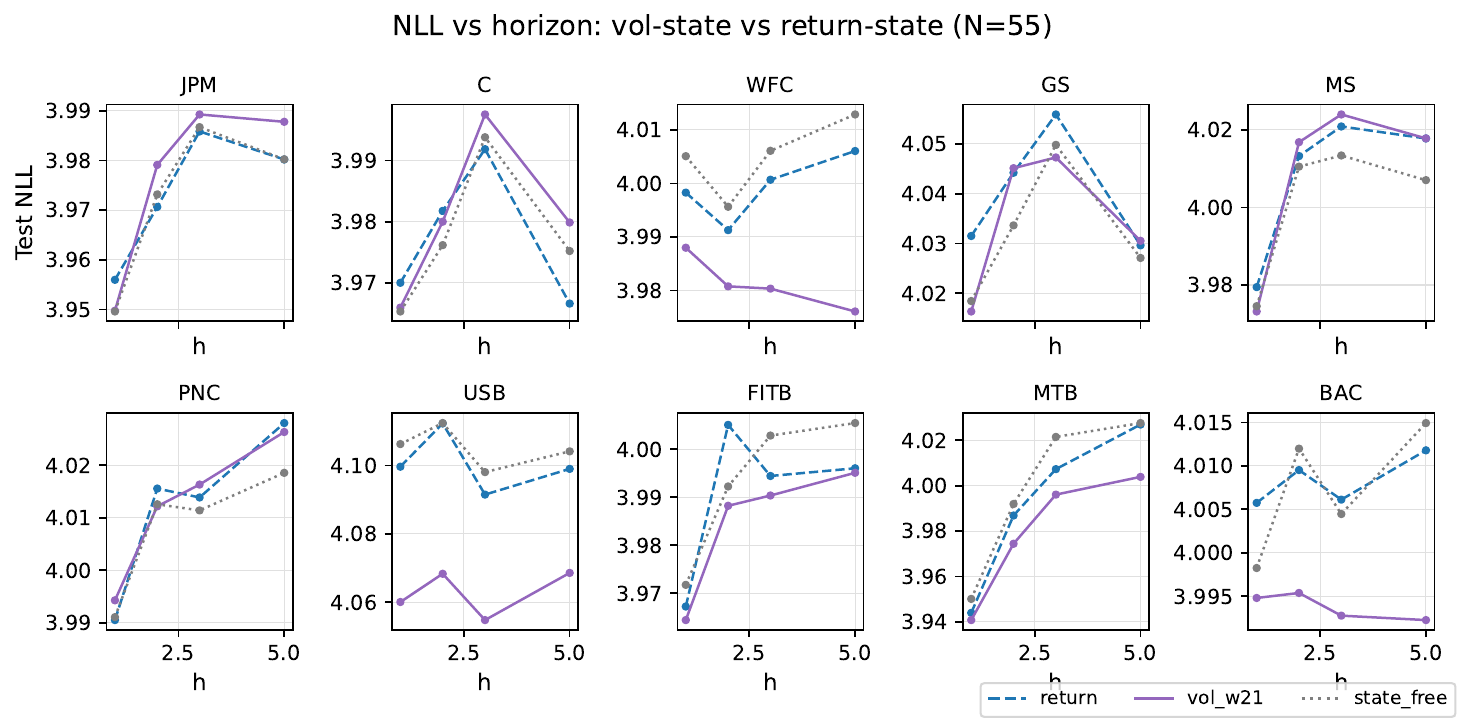}
    \caption{Test NLL vs.\ horizon $h$ for
    \texttt{state\_cond\_return}, \texttt{vol\_w21}, and
    \texttt{state\_free} at $N=55$, per asset. The vol-state
    advantage is largest at $h=1$ and narrows at longer horizons,
    with curves becoming near-overlapping for several assets at
    $h \geq 2$.}
    \label{fig:nllgrid}
\end{figure}

\paragraph{Behavior across horizons.}
Table~\ref{tab:nll_by_horizon} reports pooled NLL across horizons
$h \in \{1, 2, 3, 5\}$. The advantage of \texttt{vol\_w21} over
\texttt{state\_cond\_return} is largest at $h=1$ ($0.0095$ nats)
and narrows at longer horizons ($0.0079$ at $h=3$, $0.0084$ at
$h=5$). Figure~\ref{fig:nllgrid} shows the same pattern broken
out per asset: vol-state and return-state NLL curves become
near-overlapping for several assets at $h \geq 2$. This pattern
is consistent with the structural role of the vol state:
$\sigma_t$ encodes the regime as of time $t$, but as the
prediction horizon extends, new information realized between
$t$ and $t+h$ is not reflected in the conditioning state,
diluting its informational advantage. Recovering the
short-horizon edge at longer horizons would likely require
online updating of the state or higher-order conditioning.

Among the vol-state windows, $w=21$ dominates consistently at
every combination of $N$ and $h$, with the gap over
\texttt{vol\_w10} widening at higher resolutions. The shortest
window (\texttt{vol\_w5}) occasionally underperforms
\texttt{state\_cond\_return}, confirming that a window too
short to capture the volatility clustering timescale provides
little benefit over return-state conditioning.

\begin{table}[h]
\centering
\caption{Mean test NLL by horizon ($N=55$, pooled over 10 assets and 3 seeds, \texttt{s}=\texttt{state}). The maximum entropy level at $N=55$ is $\log 55 \approx 4.01$ nats.}
\label{tab:nll_by_horizon}
\begin{tabular}{lcccc}
\hline
\textbf{Model} & $h=1$ & $h=2$ & $h=3$ & $h=5$ \\
\hline
\texttt{s\_cond\_vol\_w21} & 3.9847 & 4.0040 & 4.0089 & 4.0078 \\
\texttt{s\_cond\_vol\_w10} & 3.9864 & 4.0057 & 4.0133 & 4.0090 \\
\texttt{s\_free}           & 3.9930 & 4.0110 & 4.0188 & 4.0173 \\
\texttt{s\_cond\_macro}    & 3.9938 & 4.0091 & 4.0185 & 4.0126 \\
\texttt{s\_cond\_return}   & 3.9942 & 4.0130 & 4.0168 & 4.0162 \\
\texttt{s\_cond\_vol\_w5}  & 3.9955 & 4.0101 & 4.0181 & 4.0164 \\
\hline
\end{tabular}
\end{table}

\subsection{Internal Consistency of the Markov Structure}
\label{sec:results_consistency}

A predictive gain alone does not establish that vol-state conditioning produces a better-behaved Markov structure. The explicit transition matrix lets us probe this directly through the Chapman-Kolmogorov (CK) identity: how well does the direct $h$-step prediction $\widehat{A}_t^{(h)}$ agree with the composition of $h$ one-step predictions $\widehat{A}_t^{(1:h)}$? A perfect first-order Markov system would show zero discrepancy. In real data, smaller discrepancy means the local Markov approximation holds more cleanly.

Across the full 24-year test set at $h=5$, the vol-state model achieves a grand mean CK KL divergence of $0.0620$, compared to $0.0657$ for the return-state model; a $5.6\%$ reduction in structural discrepancy. The pattern is consistent across the portfolio: in 8 of 10 assets, the return-state model exhibits higher CK discrepancy. BAC is the largest gap, with a return-state CK error of $0.089$ versus $0.076$ for the vol-state. The volatility-state framework is therefore not just more accurate predictively but also more mathematically coherent: its multi-step predictions are closer to what a true first-order Markov chain would produce.

We also examined the Dobrushin coefficient, which measures the maximum row separation in total variation and bounds the operator's contraction rate. The vol-state model exhibits higher Dobrushin in 4 of 10 assets, indicating no systematic difference between the two conditioning schemes. Both produce contractive operators with comparable mixing properties; the vol-state advantage lies in the Markov structure itself, not in matrix contraction behavior.

\subsection{What the Explicit Matrix Reveals}
\label{sec:results_geometry}

The Markov structure being more internally consistent is one benefit of the explicit matrix output. The other is that we can apply standard diagnostics (row entropy, pairwise correlations across assets) directly to the learned matrices and read off structural properties of the underlying dynamics that a black-box predictor would obscure.

\begin{figure}[t]
    \centering
    \includegraphics[width=1\linewidth]{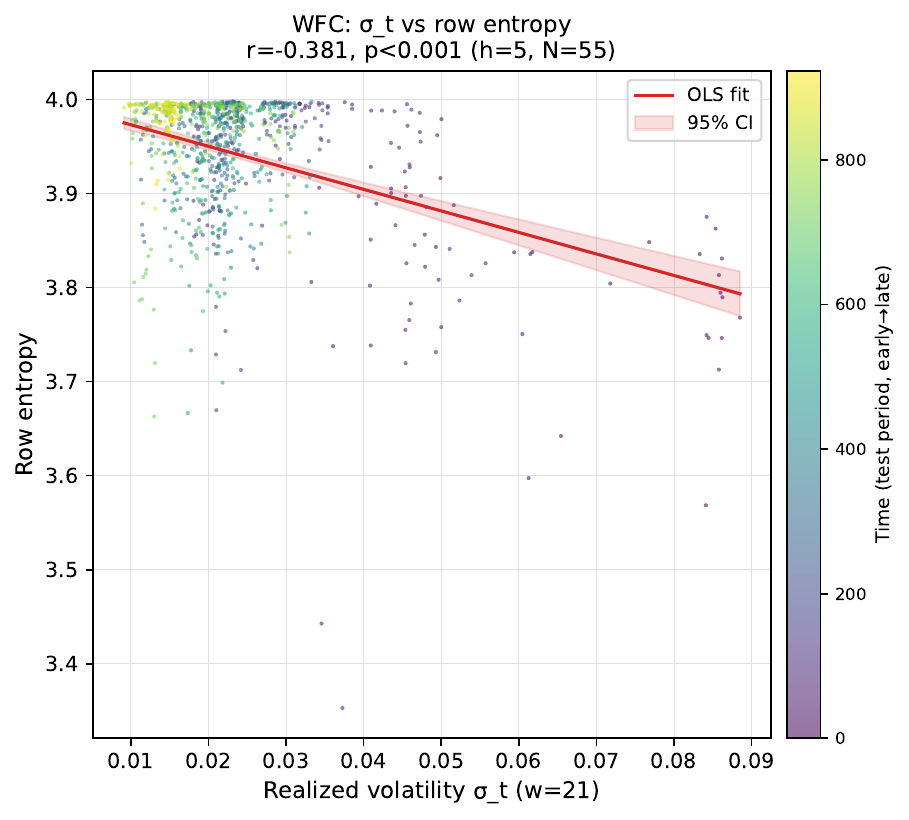}
    \caption{Realized volatility $\sigma_t^{(10)}$ vs.\ operator
    row entropy $H(\widehat{A}_t)$ over the test period (WFC,
    $h=5$, $N=55$). Each point is one trading day, coloured by
    time. The Pearson correlation is $r = -0.381$
    ($p = 2.77 \times 10^{-33}$). Across all 54 evaluated
    combinations of asset, window, and horizon, the correlation
    is negative without exception and statistically significant
    in 42 of 54.}
    \label{fig:scatter}
\end{figure}

\paragraph{High volatility homogenizes the matrix.} Figure~\ref{fig:scatter} plots realized volatility $\sigma_t^{(10)}$ against operator row entropy $H(\widehat{A}_t)$ for WFC over the test period ($h=5$, $N=55$). The Pearson correlation is $r = -0.381$
($p = 2.77 \times 10^{-33}$). This is the strongest individual relationship we observe, but the pattern is universal: across all 54 combinations of asset, window length, and horizon evaluated, the correlation is negative without exception, with 42 of 54 reaching statistical significance ($p < 0.05$) and a mean $r = -0.198$ pooled across all combinations. High volatility consistently produces more concentrated transition matrices: when $\sigma_t$ is high, the network learns that high-vol regimes drive transition rows toward similar, concentrated distributions regardless of the current state. Row entropy drops because the matrix loses its differentiation across rows, the system behaves as if the current state matters less under stress.

\begin{figure}[t]
    \centering
    \includegraphics[width=1\linewidth]{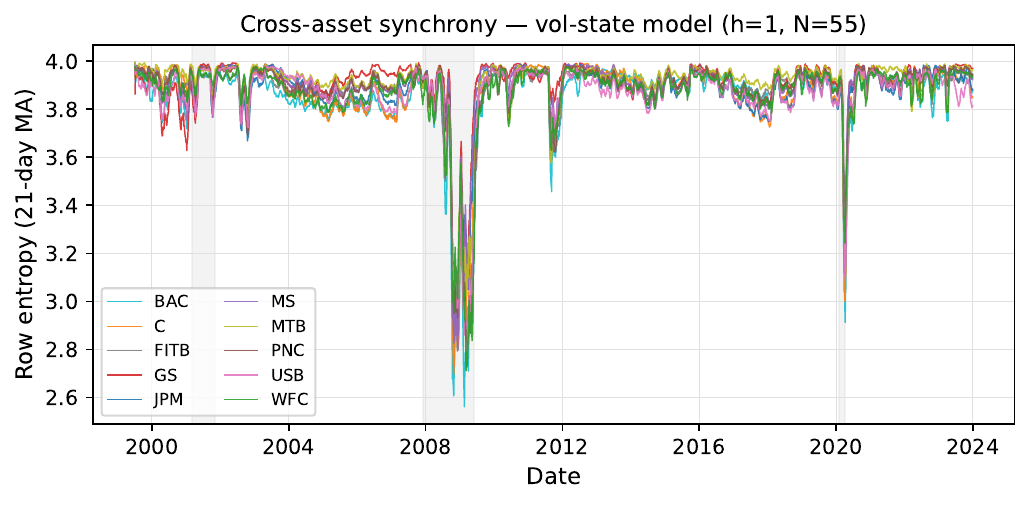}
    \caption{Row entropy $H(\widehat{A}_t)$ (21-day moving
    average) for all ten assets over the full sample
    (\texttt{vol\_w21}, $h=1$, $N=55$). Series are highly
    synchronised (mean pairwise Pearson $= 0.646$). Shaded
    regions indicate NBER recession periods: the dot-com bust
    (2001), the GFC (2007-2009), and the COVID-19 shock (2020).
    Coherent entropy dips are visible during each stress window.}
    \label{fig:crossasset}
\end{figure}

\paragraph{Cross-asset behavior during stress.}
Computing the same row entropy series for each asset and comparing across the portfolio provides a consistency check on the framework. Figure~\ref{fig:crossasset} shows the 21-day moving average of $H(\widehat{A}_t)$ for all ten assets over the full sample from June 1999. The mean pairwise Pearson correlation across the ten series is $0.646$, and the three NBER recession windows in the sample (the dot-com bust of 2001, the global financial crisis of 2007-2009, and the COVID-19 shock of 2020) appear as synchronised entropy dips across all ten assets. Given that the ten US bank stocks share substantial macro and credit exposures, this synchrony is expected; we report it as a sanity check that the framework correctly captures common factor structure rather than fitting asset-specific noise. A heterogeneous cross-asset portfolio would be needed to assess whether the synchrony reflects a deeper property of the operator framework itself.

\FloatBarrier

\section{Discussion}
\label{sec:discussion}

\subsection{Volatility as a Markovianizing State Variable}
\label{sec:disc_markovianization}

The central empirical finding of this paper is that conditioning
on realized volatility produces a more internally consistent
Markov structure than conditioning on the return level. The
$5.6\%$ reduction in CK discrepancy under vol-state conditioning
is not large in absolute terms, but its direction is the relevant
signal: across 8 of 10 assets, the composed one-step matrices
agree more closely with the directly trained multi-step matrix
when the state variable is volatility rather than the return.
The vol-state framework is closer to behaving like a true
first-order Markov chain.

This makes structural sense. The first-order Markov assumption
requires that the current state $X_t$ summarize everything
relevant about the past for predicting the future. Daily returns
do not satisfy this: under weak-form efficiency, $r_t$ is close
to serially uncorrelated, so it carries almost no information
about the regime the system is currently in or where it is
heading. Conditioning on $r_t$ therefore leaves significant unexplained patterns. Realized volatility is the opposite. Volatility clustering means $\sigma_t$ aggregates
recent realized variance into a persistent quantity that captures
much of the relevant state of the system. Conditioning on
$\sigma_t$ moves more of the temporal dependence into the
state itself, leaving less residual non-Markovian behavior for
the matrix to absorb. The CK discrepancy result is the
empirical fingerprint of this: when the state is well-chosen,
the Markov structure that the framework imposes is approximately
satisfied by the data; when the state is poorly chosen, the
framework still produces a valid stochastic matrix at every
timestep, but its internal consistency suffers.

\subsection{What the Explicit Matrix Buys}
\label{sec:disc_explicit_matrix}

The CK comparison would not be possible without the explicit
matrix output. A model that produces only a next-step
distribution, or that compresses dynamics into a latent recurrent
state, has no equivalent of $\widehat{A}_t^{(h)}$ to compare
against $\widehat{A}_t^{(1:h)}$; the question of internal
consistency cannot even be posed. The same is true for the
volatility-entropy relationship. Computing row entropy at every
timestep and correlating it with realized volatility is only
meaningful when each timestep produces an inspectable
$N \times N$ stochastic matrix on a fixed state space. The
universal negative correlation we observe ($r < 0$ in 54 of 54
configurations, significant in 42 of 54) is therefore not just
a statement about the data but about what the framework lets us
ask: it makes the structural relationship between market stress
and the geometry of transition probabilities directly observable.

This is the broader argument the paper makes for structured
probabilistic outputs in time series modeling. A neural network
that outputs a probability distribution can be evaluated for
predictive accuracy; one that outputs a stochastic matrix on a
shared state space can additionally be evaluated for internal
mathematical consistency, contraction behavior, and structural
properties of the implied dynamics. The choice of output
structure determines what questions the model is capable of
answering, independently of how it is parameterized.

\subsection{Limitations and Scope}
\label{sec:disc_limitations}

Three limitations are worth noting. First, the predictive gains
are modest in absolute terms ($0.0095$ nats pooled over
return-state at $h=1$) and narrow further at longer horizons.
This is expected at daily equity frequency given the well-known
limits of return predictability \cite{cont2001empirical}, and
the paper's central claim concerns the structural properties of
the learned matrices rather than predictive performance. The
predictive results are reported to confirm that the vol-state
framework is at least as good a forecaster as the return-state
alternative, not to argue for it primarily on those grounds.

Second, the cross-asset synchrony observed in
Section~\ref{sec:results_geometry} should be read as a
consistency check rather than a finding. The ten assets are all
US bank stocks with substantial shared macro and credit
exposures; high pairwise correlation of their row entropy
series is largely a consequence of that selection. A
heterogeneous portfolio (mixing equities, fixed income,
commodities, and FX) would be needed to assess whether the
operator-level synchrony reveals something deeper about
cross-asset stress propagation, or whether it simply reflects
the common factor structure of the input data.

Third, USB is a clean exception to the vol-state advantage:
all conditioning schemes fail to beat the marginal baseline at
$h=1$. The pattern points to idiosyncratic dynamics at daily
frequency that none of the available state variables capture.
Per-asset feature engineering, learned state representations, or
higher-frequency data may be needed for assets where the macro
and cross-asset feature set is insufficient.

\subsection{Generalization Beyond Finance}
\label{sec:disc_generalization}

The framework is presented through a financial application, but
the design principle generalizes. Any time-series modeling
problem where (i) a discrete state space is meaningful, (ii) the
underlying process is non-stationary, and (iii) a domain-motivated
candidate state variable carries serial dependence, can in
principle benefit from the same construction: parameterize the
transition matrix with a neural network constrained to output
valid stochastic rows, condition on the persistence-bearing state
variable, and use the resulting explicit matrix to apply
classical operator diagnostics. Examples include credit-rating
migration, weather and climate transitions, biological state
sequences, and queueing systems. The contribution of this paper
is to demonstrate, in a single domain where the relevant state
variable is well-understood, that the choice of state variable is
a substantive modeling decision with measurable consequences for
how Markovian the resulting structure is.

\section{Conclusion}
\label{sec:conclusion}

We presented a framework that uses a neural network to
parameterize time-inhomogeneous Markov transition matrices,
allowing the matrices to be estimated at fine discretizations
where empirical counting fails. The framework's main design
choice is the Markov state variable, and the central claim of
the paper is that this choice has measurable structural
consequences.

Conditioning on realized volatility rather than the return level
produces a Markov structure that better satisfies the first-order
assumption it is built on: across the ten US bank stocks
evaluated, the volatility-state model achieves a $5.6\%$ lower
mean Chapman-Kolmogorov discrepancy than the return-state
alternative, with the gap holding in 8 of 10 assets. The
mechanism is direct. Daily returns are nearly serially
uncorrelated and therefore carry little information about the
current regime; realized volatility is persistent and aggregates
recent information in a way that returns do not. Moving the
serial dependence into the state itself is what allows the
resulting Markov chain to be approximately self-consistent across
horizons. Vol-state conditioning also wins on held-out likelihood
in 9 of 10 assets at $h=1$, but the structural result is the
load-bearing one.

The framework's explicit matrix output is what makes these
structural claims testable. The CK comparison requires being able
to compose one-step matrices and compare against a directly
trained multi-step matrix; the universal negative correlation between volatility and row entropy ($r < 0$ in 54 of 54 configurations) requires being able to compute row entropy at every timestep. Neither diagnostic exists for a model that represents dynamics through latent recurrent states or single predictive distributions. The choice of output structure determines which questions a model is capable of answering.

These ideas: choosing a state variable that has a strong autocorrelation, parameterizing the transition matrix explicitly, and using the resulting matrix to verify the structural assumptions of the model, apply beyond financial time series to any non-stationary discrete-state process where the standard estimation routes break down.

\bibliographystyle{plainnat}
\bibliography{references}

\end{document}